\documentclass{article}
\usepackage{ijcai18}

\usepackage{xcolor}
\usepackage{soul}
\usepackage[utf8]{inputenc}
\usepackage[small]{caption}
\usepackage{latexsym}
\usepackage{amsmath}
\usepackage{amssymb}
\usepackage{graphicx}
\usepackage{todonotes}
\usepackage{caption}
\usepackage[mathscr]{euscript}
\usepackage[T1]{fontenc}
\usepackage{hyperref}
\usepackage{booktabs} 
\usepackage{times}    % Integrate Times for here
\usepackage{cite}     % Automatically ordered citations
\usepackage{xspace}
\usepackage{graphicx}
\usepackage{algorithmic}
\usepackage{algorithm}
\usepackage{multirow}
\usepackage{indentfirst}
\usepackage[numbers,sort&compress]{natbib}

\usepackage{url}
\usepackage[inline]{enumitem}

\title{
Multimodal Interaction Modeling via
       Self-Supervised Multi-Task Learning for \\
       Review Helpfulness Prediction
}

\iftrue
\author{
HongLin Gong$^1$, 
Mengzhao Jia$^2$, 
Liqiang Jing$^3$
\\ 
$^1$ Sun Yat-sen University,Guangzhou, China, {\tt\small gonghlin@mail2.sysu.edu.cn}\\
$^2$ University of Notre Dame,Notre Dame, USA, {\tt\small mjia2@nd.edu}\\
$^3$ University of Texas at Dallas, Richardson, USA, {\tt\small jingliqiang6@gmail.com}  \\
}
\fi

\begin{document}

\maketitle

\begin{abstract}
In line with the latest research, the task of identifying helpful reviews from a vast pool of user-generated textual and visual data has become a prominent area of study. Effective modal representations are expected to possess two key attributes: consistency and differentiation. Current methods designed for Multimodal Review Helpfulness Prediction (MRHP) face limitations in capturing distinctive information due to their reliance on uniform multimodal annotation. The process of adding varied multimodal annotations is not only time-consuming but also labor-intensive. To tackle these challenges, we propose an auto-generated scheme based on multi-task learning to generate pseudo labels. This approach allows us to simultaneously train for the global multimodal interaction task and the separate cross-modal interaction subtasks, enabling us to learn and leverage both consistency and differentiation effectively. Subsequently, experimental results validate the effectiveness of pseudo labels, and our approach surpasses previous textual and multimodal baseline models on two widely accessible benchmark datasets, providing a solution to the MRHP problem.
\end{abstract}

\section{Introduction}
\label{sec:intro}

The exponential growth of information has resulted in a disparity in the quality of user-generated reviews, inundating people with an ever-expanding volume of noisy data while shopping online. Consequently, intelligently filtering out low-quality reviews and highlighting the most helpful ones at the top position holds advantages for both users and suppliers. This trend has recently sparked a surge of interest in addressing the Review Helpfulness Prediction (RHP) problem.

In the early stages of RHP research, efforts primarily revolved around plain textual datasets. One approach to solving this challenge was through machine learning. However, this method proved to be labor-intensive and struggled with processing irregular raw data. Subsequently, scholars turned to deep neural networks to extract valuable information from the original data. Yet, these approaches faced limitations since reviews were not solely presented in textual form but also in other modalities. Consequently, attempts were made to incorporate the visual modality. Nevertheless, these endeavors fell short as they did not distinguish and explicitly model the intricate correlations among multimodalities.

\begin{figure}[ht]
    \centering
    \includegraphics[width=\linewidth]{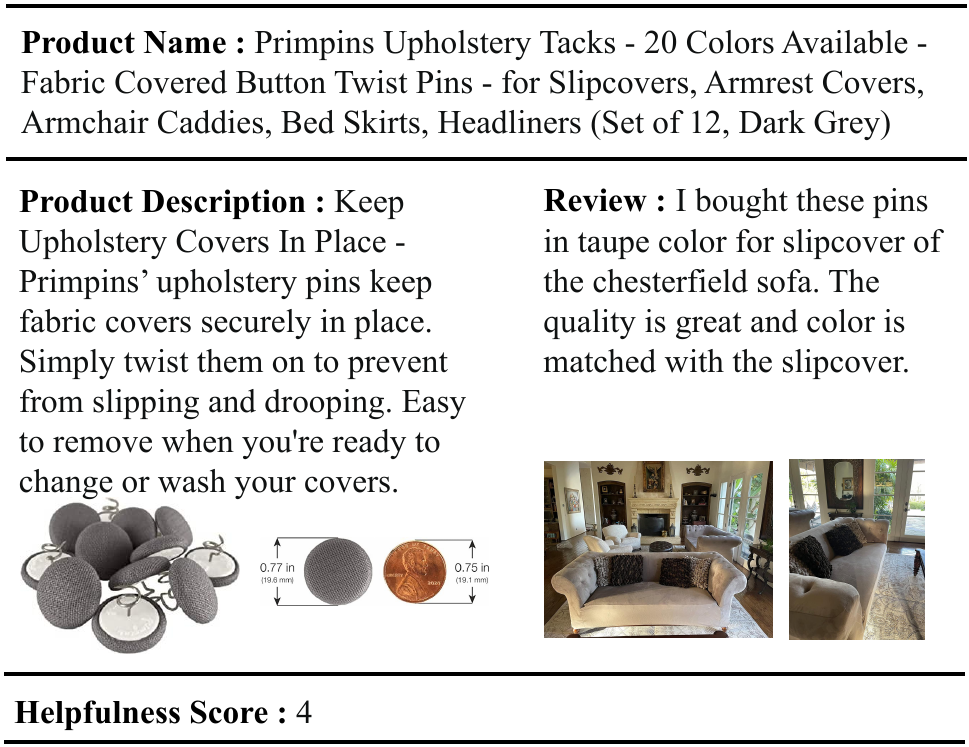}
    \caption{Illustration of the consistent and different correlations between the textual and visual modalities.}
      \label{fig:label1}
\end{figure} 

 In an ideal scenario, the combination of textual and visual modalities should jointly represent the same item, thus sharing a certain degree of consistency. However, reviews often introduce variations in the connotative meanings conveyed by these two modalities. As illustrated in Figure \ref{fig:label1}, a customer purchased pins to repair the edge of their sofa, but instead of capturing photos of the pins, they only presented an image of the neatly restored sofa after the pins were installed. In this example, the images of the product and the review appear unrelated, which typically leads to a prediction of unhelpfulness. Nevertheless, by considering the accompanying review text, we contend that the review should be deemed helpful and deserving of a high score.

Previous research~\cite{ref15,ref17} has predominantly concentrated on achieving consistent representations between textual and visual modalities, often neglecting the exploration of their differences. To fully leverage these two aspects, we introduce a \textbf{M}ultimodal Interaction \textbf{M}odel via \textbf{S}elf-\textbf{S}upervised Multi-task Learning, abbreviated as \textbf{MM-SS} in this paper. Our approach begins by uncovering multiple perspective interactions within feature vectors across various domains (e.g., products and reviews) and modalities (e.g., text and vision). Subsequently, we fuse these interactions to create a comprehensive representation that captures global consistency. Furthermore, we decompose the global representation into multiple independent subtasks to refine differences, a significant innovation in our approach. To achieve this, we generate pseudo labels through a self-supervised learning mechanism. Our design is grounded in two fundamental intuitions. Firstly, modal representations exhibit strong correlations with labels. Secondly, pseudo labels can be derived from multimodal manual annotations. This motivates us to automatically generate pseudo labels by calculating the relevance-based distance between labels. In essence, we concatenate all separate interactions into a unified representation to capture their 'consistent' aspects. Simultaneously, we establish independent subtasks using these separate interactions to accentuate their 'distinct' features. This enables them to collectively influence the final consensus representation and enhance the overall performance, as detailed in Section \ref{section:D}.

In summary, this study makes the following contributions:
\begin{itemize}
\item Simultaneous Utilization of ``Consistency'' and ``Difference'': We introduce a self-supervised multi-task learning scheme for the Multimodal Review Helpfulness Prediction (MRHP) task, effectively harnessing the ``Consistency'' and ``Difference'' of modal representations.
\item Automated Pseudo Label Generation: This research presents a label generation strategy that enables the automatic creation of pseudo labels, eliminating the need for additional manual labeling efforts.
\item Superior Performance: Extensive experiments conducted on two datasets for MRHP problems demonstrate that our approach surpasses both plain textual and multimodal benchmarks in terms of performance.
\end{itemize}

\section{Related Work}
\label{sec:related work}

\noindent\textbf{Review Helpfulness Prediction.}
Previous endeavors in review helpfulness prediction primarily concentrated on text-only approaches, involving the extraction of a wide array of hand-crafted features, such as structural features~\cite{ref7}, lexical features~\cite{ref8}, semantic features~\cite{ref2}, emotional features~\cite{ref1}, and argument features~\cite{ref3}.
However, these approaches heavily relied on feature engineering, a process known for its time-consuming and labor-intensive nature. With the remarkable growth in available computational resources, the research focus has shifted towards the utilization of deep neural networks.
In practice, the visual modality (i.e., images) has emerged as an increasingly crucial source of essential information. As a result, Liu et al.~\cite{ref15} introduced a new task in Multimodal Review Helpfulness Prediction, incorporating both textual and visual modalities. Subsequently, Han et al.~\cite{ref16} introduced a probe-based approach to uncover the inherent relationships within data from various domains and modalities.
Concurrently, Nguyen et al.~\cite{ref17} proposed a cross-modal contrast learning model to extract mutual information from input modalities.

Distinguishing itself from the aforementioned methods, our paper employs a self-supervised multi-task learning strategy that combines global multimodal interaction with separate cross-modal interaction. This approach aims to strike a balance between consistency and diversity.

\begin{figure*}[ht]
\centering
\includegraphics[width=0.9\textwidth]{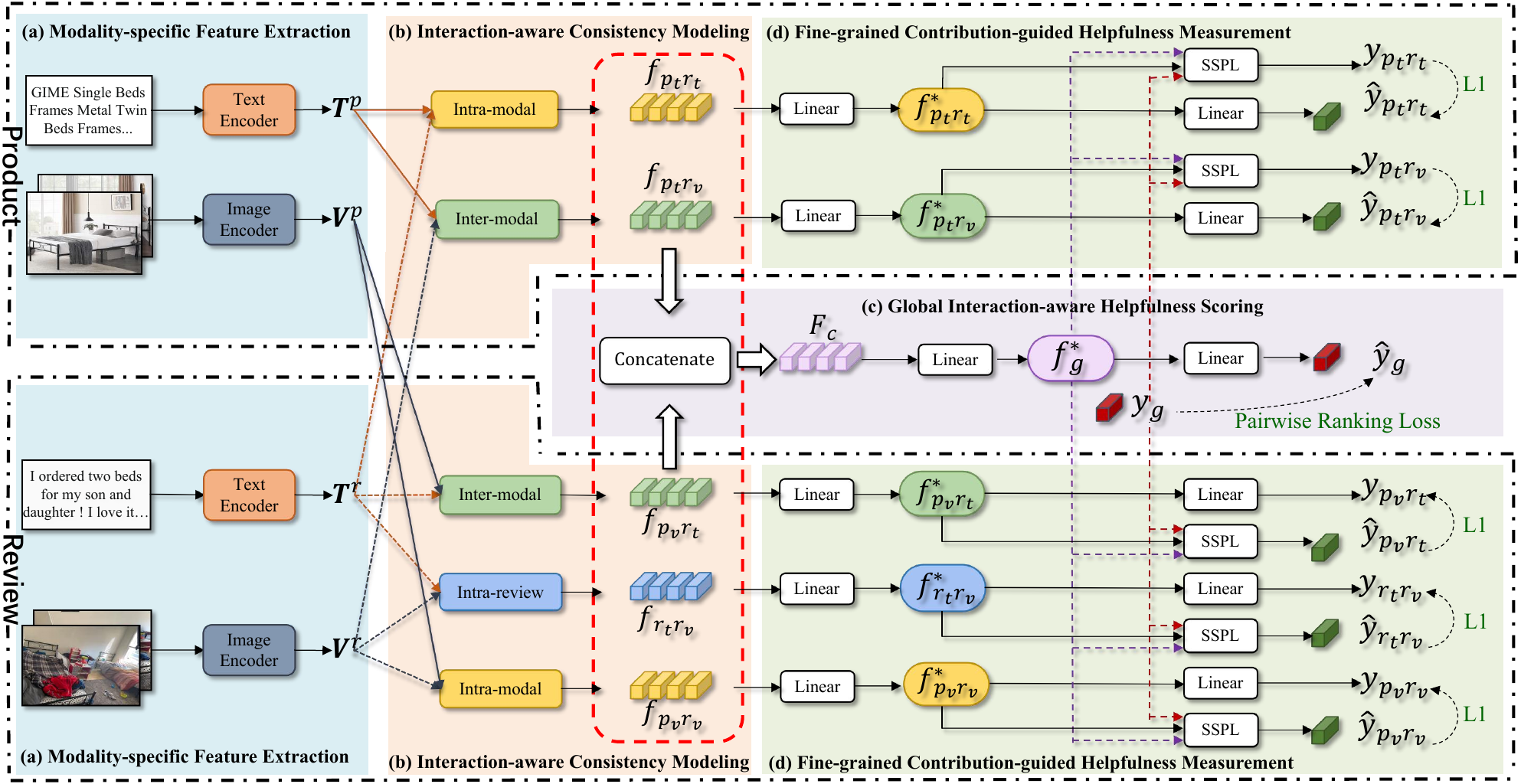}
\caption{Illustration of the proposed MM-SS scheme. It consists of four modules: (a) \textit{Modality-specific Feature Extration}, (b) \textit{Interaction-aware Consistency Modeling}, (c) \textit{Global Interaction-aware Helpfulness Scoring}, and (d) \textit{Fine-grained Contribution-guided Helpfulness Measurement}. } 
\label{fig:label2}
\end{figure*}
\vspace{5pt}

\noindent\textbf{Multi-Task Learning.} 
Multi-task learning is employed to enhance the model's generalization performance by establishing shared network structures and effectively balancing the training of multiple tasks~\cite{ref38,vkg2t,DBLP:journals/tois/JingSLZZN24,DBLP:journals/tkde/SunJWSCN23}. An exemplary network structure in a model not only consolidates common knowledge across diverse tasks but also guards against task interference. Two distinct approaches, namely soft parameter sharing~\cite{ref43,ref44} and hard parameter sharing~\cite{ref41,ref42}, are adopted, contingent on the extent of parameter sharing among tasks. In soft parameter sharing, independent parameters are assigned to individual tasks, with shared knowledge facilitating the optimization of additional constrained functions. On the other hand, hard parameter sharing involves sharing parameters in the core of the model, while task-specific parameters near the output remain independent.

Another noteworthy aspect is the allocation of suitable weights to different tasks by defining a comprehensive loss function. This weight assignment is crucial for balancing the training of each task to yield favorable outcomes. To address this concern, Chen et al.~\cite{ref45} introduced a gradient normalization technique to automatically adjust the weight of the multi-task loss function. Additionally, Kendall et al.~\cite{ref47} proposed an automatic weighting mechanism that takes into account the homoscedastic uncertainty of each task. Furthermore, Sener et al.~\cite{ref46} treated multi-task learning as a multi-objective optimization problem, leading to the determination of optimal weights for different tasks through the Pareto Optimal Solution~\cite{ref48}.

In our work, we executed separate subtasks to support the global multimodal representation. For the sake of simplicity and effective handling of closely related or a limited number of tasks, we employed the hard parameter-sharing strategy. This approach helps significantly reduce the risk of overfitting. Consequently, we adopted the hard parameter-sharing strategy to formulate an adaptive weight-adjustment method, which ensures a balanced learning process across different tasks.

\section{Problem Formulation}
\label{sec:formulation}
Suppose there is a product item ${P}$, which consists of a textual description $D=(d_1,\cdots,d_{|D|})$ and a set of product images denoted as $\mathcal{V}=\{v_{1},\cdots, v_{|\mathcal{V}|}\}$. In this case, $d_i$ represents the $i$-token of a sum of $|D|$ description tokens and $v_i$ refers to the $i$-th image of a total number of $|\mathcal{V}|$ images.
Meanwhile, the product item contains a set of $N$ corresponding reviews denoted as $\mathcal{R} = \{{R}^{1},\cdots,{R}^{j},\cdots,{R}^{N}\}$, where the $j$-th review ${R}^{j}$ is comprised of a user-generated text $T^{j}=(t_1^{j},\cdots,t_{|T^{j}|}^{j})$ and a set of images $\mathcal{K}^{j} =\{k^{j}_1,\cdots, k^{j}_{|\mathcal{K}^{j}|}\}$. $T^{j}$ and $\mathcal{K}^{j}$ contain $|T^{j}|$ tokens and $|\mathcal{K}^{j}|$ images, respectively.
In addition, the $j$-th review ${R}^j$ is annotated with a helpfulness score $y_j\in\{0,1,2,3,4\}$.

In the multimodal RHP task, our aim is to design a model $\mathcal{F}$ that can be used to predict the helpfulness score for the review by leveraging the textual and visual modalities of both the product item and the corresponding reviews,

\begin{equation}
\hat{y}_j=\mathcal{F}(P, {R}^j|\Theta),
\end{equation}
where $\Theta$ represents all the to-be-learned parameters of $\mathcal{F}$, and $\hat{y}_j$ refers to the predict helpfulness score of the $j$-th review ${R}^j$ for the product $P$. 
For simplicity, the index is temporarily omitted (\textit{i.e.}, the superscript $j$).

\section{METHODOLOGY}
\label{sec:method}
This section elaborates on the proposed \textbf{M}ultimodal Interaction \textbf{M}odeling via \textbf{S}elf-\textbf{S}upervised Multi-task Learning scheme \textbf{MM-SS}, as shown in Figure \ref{fig:label2}.

\subsection{Modality-specific Feature Extraction}
\label{section:A}
This study explores two modality-specific feature extractors to separately deal with the textual (product textual description $D$ and review text $T$) and visual (product image set $\mathcal{V}$ and review image set $\mathcal{K}$) inputs, respectively.

\noindent\textbf{Textual Feature Extraction.}
Product description $D$ and review text $T$ are both in the form of natural language. After adding a special token $\mathrm{[CLS]}$ as prefix, product and review tokens are fed into a pre-trained 12-layer BERT model~\cite{ref27} to determine the corresponding textual features, which has shown success in various language processing tasks~\cite{debias,DBLP:conf/mm/SunN0JWN23,DBLP:conf/mm/SunWJCSN22}, as follows,

\begin{equation}
 \left\{
\begin{aligned}
\textbf{T}^{p} &= \operatorname{BERT}(\mathrm{CLS},d_1,d_2,\cdots,d_{|D|}), \\
\textbf{T}^{r} &=\operatorname{BERT}(\mathrm{CLS},t_1,t_2,\cdots,t_{|T|}),
\end{aligned}
\right.
\end{equation}
where $\textbf{T}^{p}\in\mathbb{R}^{(|D|+1)\times d_{t}}$ and $\textbf{T}^{r}\in\mathbb{R}^{(|T|+1)\times d_{t}}$ stand for the product description feature and the review text feature, respectively. The dimension of the features is denoted as $d_t$.

\noindent\textbf{Visual Feature Extraction.}
In the context of visual signals, we employ a pre-trained Faster R-CNN model as outlined in~\cite{ref29} to identify candidate boxes that warrant special attention, which has been applied in various tasks~\cite{DBLP:conf/acl/JingSOJN23,DBLP:conf/aaai/QiaoJSCZN23}. For each image denoted as $v_{|i|}$ within the product image set $\mathcal{V}$, we extract the the Regions of Interest (RoI) features from the last hidden layer just prior to the classifier, which are represented as $\mathbf{V}_i$. This process is extended to the review image set, and for the the $|i|$-th review image, we denote its RoI feature as $\mathbf{K}_i$. Subsequently, we combine all the extracted features using a concatenation operation from both the product and review image sets.
Following this, motivated by the existing work~\cite{DBLP:journals/ijautcomp/JingLXYSS23,DBLP:conf/sigir/LinJSLSN23}, we apply a Self-Attention mechanism~\cite{ref30} to the amalgamated features. These features are then projected into a $d_v$-dimensional semantic space that enables non-local comprehension. This transformation allows us to represent the visual semantic characteristics of the product image sets as $\textbf{V}^p$, and the visual semantic characteristics of the review image sets as $\textbf{V}^r$, respectively.

\subsection{Interaction-aware Consistency Modeling}
\label{BB}
In fact, the helpfulness of a review is influenced by two factors. Firstly, the inter-modal consistency of the multimodal review determines whether the review is of high quality. Secondly, the alignment within the review-product pair indicates whether the review can provide the information closely related to the particular product.

\noindent\textbf{Review-only Inter-modal Consistency.}
We propose to model the inter-modal semantic relationship of the multimodal reviews. So we explore the interaction between the textual (\textit{i.e.}, $\mathbf{T}^{r}$) and visual (\textit{i.e.}, $\mathbf{V}^{r}$) features. We adopt a Self-Attention~\cite{ref30} layer to integrate the features from different modalities,
\begin{equation} 
\mathbf{C}_{r}=\operatorname{Self-ATT}([ \mathbf{T}^{r};\mathbf{V}^{r}]),
\end{equation}
where $\mathbf{C}^{r}$ denotes the multimodal representation of the review, and $\mathbf{C}^{r}$ passes through a Soft Pooling layer~\cite{ref32} to determine the final inter-modal representation $\mathbf{f}_{r_{t}r_{v}}$.

\noindent\textbf{Multisource Review-Product Alignment.}
With regard to the alignment between review and product, this study takes into account both the inter- and intra-modal semantic relationships. In total, it explores four types of relationship between review and product feature (\textit{i.e.}, $\mathbf{T}^{p}$, $\mathbf{T}^{r}$, $\mathbf{V}^{p}$, and $\mathbf{V}^{r}$). 

\textbf{Intra-modal Interaction}.
Firstly, we give consideration to the semantic relations within the same modality: (i) product text - review text and (ii) product vision - review vision. We adopt a Self-Attention layer following previous study~\cite{ref17} to derive the intra-modal representation, which is then fed into a $\operatorname{SoftPooling}$ layer for the final intra-modal representations,
\begin{equation} \left\{
\begin{aligned}
\mathbf{f}_{p_{t}r_{t}}&=\operatorname{SoftPool}(\operatorname{Self-ATT}([\textbf{T}^{p};\textbf{T}^{r}])),\\
\mathbf{f}_{p_{v}r_{v}}&=\operatorname{SoftPool}(\operatorname{Self-ATT}([\textbf{V}^{p};\textbf{V}^{r}])),
\end{aligned}
\right.
\end{equation}
where $\{p,r\}$ represents the review and product description fields and $\{t,v\}$ refers to the textual and visual modal types.

\textbf{Inter-modal Interaction}. 
Secondly, we focus on cross-modal interactions and extract two intrinsic factors between textual and visual modals: (i) product text - review vision and (ii) product vision - review text,

\begin{equation} \left\{
\begin{aligned}
\mathbf{f}_{p_{t}r_{v}}&=\operatorname{SoftPool}(\operatorname{Self-ATT}([\textbf{T}^{p};\textbf{V}^{r}])),\\
\mathbf{f}_{p_{v}r_{t}}&=\operatorname{SoftPool}(\operatorname{Self-ATT}([\textbf{V}^{p};\textbf{T}^{r}])),
\end{aligned}
\right.
\end{equation}

At last, we obtain five types of cross-modal interactive states, \textit{i.e.},  $p_{t}r_{t},p_{v}r_{v},p_{t}r_{v},p_{v}r_{t}, r_{t}r_{v}$.

\subsection{Global Interaction-aware Helpfulness Scoring}
\label{section:C}
It is intuitive to derive an overall output for prediction by fusing these representations. We adopt the most straightforward fusion approach, \textit{i.e.}, concatenation, formally as,
\begin{equation}
    \mathbf{F}_c = [\mathbf{f}_{p_{t}r_{t}}; \mathbf{f}_{p_{v}r_{v}}; \mathbf{f}_{p_{t}r_{v}}; \mathbf{f}_{p_{v}r_{t}}; \mathbf{f}_{r_{t}r_{v}}],    
\end{equation}
where $ \mathbf{F}_c $ indicates the fused representation,  $[\cdot;\cdot;\cdot]$ denotes the concatenation operation. It is sent into a linear layer to project onto a lower-dimensional space, formally as,
\begin{equation} 
\mathbf{f_g^*}= \operatorname{GELU}(\operatorname{Linear(\mathbf{F}_c)}),
\end{equation}
where, we adopt the $\operatorname{GELU}$~\cite{ref50} activation function. 

As described in previous studies~\cite{ref15,ref16,ref17}, we treat RPH problem as a regression task, where each review is corresponding to a single helpfulness score. Empirically, the global representation $\mathbf{f_g^*}$ passes through a linear layer to to predict the helpfulness score,
\begin{equation}
\hat{y}_{g}={\textbf{W}^{g}}^Tf_{g}^*+\textbf{b}^{g},
\end{equation}
where $\textbf{W}^{g}\in\mathbb{R}^{d_{g}\times1} $.

\subsection{Fine-grained Contribution-guided Helpfulness Measurement}
\label{section:D}
Our aim is to guide above global interactive tasks in a self-supervised manner by employing separate subtask outputs. From \ref{BB}, we obtain five distinct interactive representations inspired by \cite{ref56}. These representations serve as the basis for establishing five independent subtasks for generating pseudo-labels. To mitigate dimensional variances across these subtasks, we project them into a new feature space. Subsequently, we derive separate interaction results using linear regression,
\begin{equation} \left\{
\begin{aligned}
f^{*}_{s}&=\operatorname{GELU}({\textbf{W}^{s}_{\alpha1}}^{T}f_{s}+\textbf{b}_{\alpha1}^{s}),\\
\hat{y}_{s}&=\operatorname{MLP}(f^{*}_{s}),
\end{aligned}
\right.
\end{equation}
where $s\in\{p_{t}r_{t},p_{v}r_{v},p_{t}r_{v},p_{v}r_{t}, r_{t}r_{v}\}$, $f_s^*$ denotes the separate interactive representation, $\hat{y}_s$ denotes the predictive output of the separate interaction task.

To circumvent manual labeling and direct the training of pseudo label generation subtasks, we built an \textbf{S}elf-\textbf{S}upervised \textbf{P}seudo-\textbf{Label} Generator (SSP-Label for short). SSP-Label aims to generate pseudo labels based on multimodal annotations by human and multimodal representations,
\begin{equation}{y}_{s}=\operatorname{SSP-Label}(y_{g},f^{*}_{g},f^{*}_{s}), \end{equation}
where $y_s$ denotes the pseudo label, $y_g$ denotes the multimodal manual annotation, $f_g^*$ denotes the global representation. It's worth noting that $y_s$ is not a real entity and only exists during the training phase. Therefore, we use $\hat{y}_{g}$ as the final helpfulness score.

\noindent\textbf{SSP-Label.}
Our approach is founded on two fundamental principles. Firstly, there exists a strong correlation between modal representations and labels, allowing us to infer that ${f_{g}^*} \propto {y_{g}}$, ${f_{s}^*} \propto {y_{s}}$. Secondly, pseudo labels demonstrate a significant connection with multimodal manual annotations, leading to the deduction that ${y_{s}} \propto {y_{g}}$. With the above insights, it becomes evident that $\frac {y_{s}} {y_{g}}\propto \frac {{f}_{s}^*} {{f}_{g}^*}$. Here we introduce an anchor point $\textbf{A}$, we posit that the global representation $f_g^*$ is in close proximity to $\textbf{A}$ and given the variability in modal representations, we assume that the separate representation $f_s^*$ is remote to $\textbf{A}$. The correlation between representations and the difference in distance from the representations to the anchor, allows us to deduce that the label difference is associated with this distance difference, where the distance from representation to the anchor denotes as $\chi$. Consequently, we find that $\frac {{y}_{s}} {{y}_{g}}\propto \frac {{\chi}_{s}} {{\chi}_{g}}$. With this understanding, we are equipped to calculate pseudo labels by incorporating an offset. Our method, denoted as $\operatorname{SSP\_Label}$, involves computing the offset $\delta_{gs}$ as depicted in Figure \ref{fig:label3}. The formula derivation for calculating the offset value is presented in the Appendix.

\begin{figure}[ht]
  \centering
  \includegraphics[width=\linewidth]{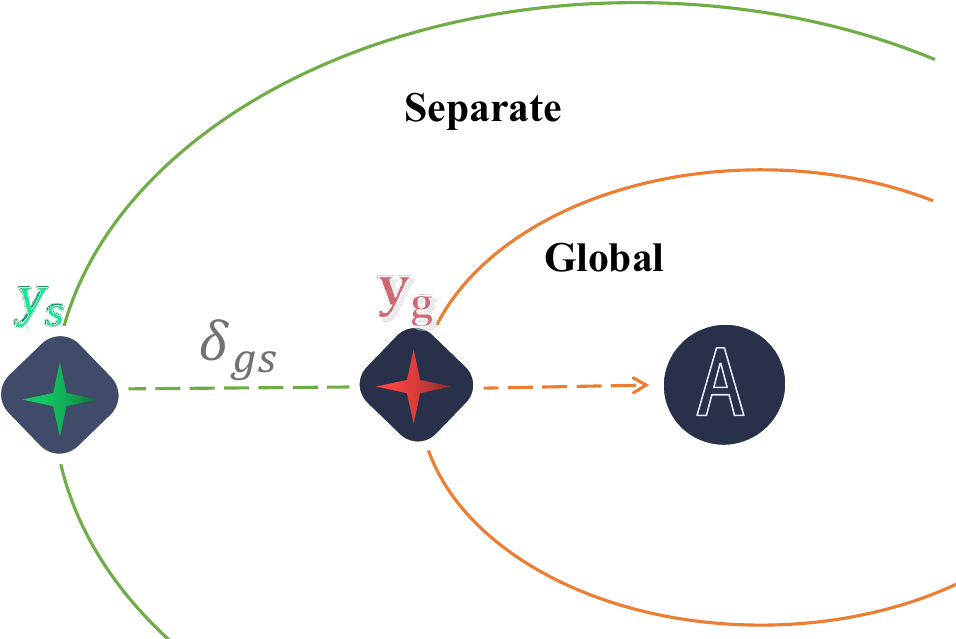}
  \caption{Pseudo label generation example, we can conclude the relationship: $y_{s}=y_{g}+\delta_{gs}$, where $\delta_{gs}$ is offset, and $s\in\{p_{t}r_{t},p_{v}r_{v},p_{t}r_{v},p_{v}r_{t}, r_{t}r_{v}\}$.}
  \label{fig:label3}
\end{figure}

\noindent\textbf{Exponentially Weighted Moving Average (EWMA)-based Update Policy.} Given the dynamic nature of feature representation, the subtask supervisions computed using Equation \ref{con:inventoryflow} in Appendix can exhibit instability. To mitigate these undesirable effects, we have devised a self-correcting update policy based on EWMA. This strategy involves adjusting current observations using historical empirical values. We posit that observations closer to the current time point exert a stronger influence on the predicted value, rendering it more representative of recent trends. 
The expression for EWMA is as follows,
\begin{equation}{y}_{s}^{(i)}=\beta \hat{y}_{s}^{(i)}+(1-\beta)y_s^{i-1}, \end{equation}
where $s\in\{p_{t}r_{t},p_{v}r_{v},p_{t}r_{v},p_{v}r_{t}, r_{t}r_{v}\}$, $y_s^i$ is the new generated pseudo-labels at the $i$-th epoch. $\hat{y}_s^{(i)}$ is the estimated value after the $i$-th epoch, the coefficient $\beta$ represents the weighting factor. However, there is a problem initializing $y_s^0=0$, as a result, the initial values are too small, and over time, this effect will gradually accumulate, resulting in large deviations. So we make changes to Equation \ref{con:inventoryflow},
\begin{equation}y_{s}^{(i)}=\left\{ 
    \begin{array}{lc}
        y_{g} & i = 1, \\
        \beta \hat{y}_s^{(i)}+(1-\beta)y_s^{(i-1)} &i>1,\\
    \end{array}
\right. \end{equation}
where $i=1,2,...n$, $\beta(0<\beta<1)$ is the weighting factor for historical measurements, and $\beta=\frac{2}{i+1}$. Because the pseudo labels are gradually accumulated over all previous training epochs, they will become stable after enough iterations (about 15 in our experiments). 
%The pseudo-labels update policy is shown in the Appendix.

For the MRHP problem, the model's parameters are updated according to the pairwise ranking loss, we calculate target task loss as follows:
\begin{equation}\mathcal{L}_{tar} =\sum\nolimits_i(0,\alpha-f(p_i,r^+)+f(p_i,r^-)),  \end{equation}
where $r^+$ and $r^-$ are random reviews in which $r^+$ possesses a higher helpfulness score than $r^-$. We calculate subtask loss as follows:
\begin{equation}\mathcal{L}_{sub} =  \sum\limits_{s}^{p_{t}r_{t},p_{v}r_{v},p_{t}r_{v},p_{v}r_{t}, r_{t}r_{v}}W_s^i \cdot  | \hat{y}_{s}^i-y_{s}^{(i)} |, \end{equation}
where  $W_s^i = tanh(|y_{s}^{(i)}-y_{g}|)$ is the weight of the $i_{th}$ sample for subtask $s$. Considering these different tasks derived from the same data set, we employ the uncertainty weighting~\cite{ref28} of the loss optimization, which uses the homoscedastic uncertainty to adjust the weight coefficient and balance the single-task loss. The model's uncertainty, independent of input data, is not a model output, but a quantity that remains constant for all input data and varies from different tasks. The optimization procedure utilizes homoscedasticity uncertainty to maximize Gaussian likelihood estimates,
\begin{equation}\begin{aligned}\mathcal{L}_{sub} =&\frac{1}{2\sigma_1^2}\mathcal{L}_{1}(W_1)+\frac{1}{2\sigma_2^2}\mathcal{L}_{2}(W_2)+\frac{1}{2\sigma_3^2}\mathcal{L}_{3}(W_3)+\\ &\frac{1}{2\sigma_4^2}\mathcal{L}_{4}(W_4) + \frac{1}{2\sigma_5^2}\mathcal{L}_{5}(W_5) + log\sigma1\sigma2\sigma3\sigma4\sigma5,\end{aligned}\end{equation}
where $\mathcal{L}_{1},\mathcal{L}_{2},\mathcal{L}_{3},\mathcal{L}_{4}$, and $\mathcal{L}_{5}$ represent the losses of the five subtasks, respectively. $\sigma_1,\sigma_2,\sigma_3,\sigma_4$ and $\sigma_5$ are the standard deviations of the Gaussian distribution, and the model's noise as well, which can balance the task-specific loss. Lastly, We jointly combine the subtasks' goal with the ranking objective of the MRHP problem to train our model,
\begin{equation}\begin{aligned}\mathcal{L} =\mathcal{L}_{tar}+ \mathcal{L}_{sub}. \end{aligned} \end{equation}
\section{EXPERIMENT}
\label{sec:experi}
% In this section, we conduct experiments on two real-world datasets by answering the following research questions.
% \begin{itemize}
%     \item \textbf{RQ1: }Does MM-SS outperform state-of-the-art baselines?
%     \item \textbf{RQ2: }How does each module affect MM-SS?
%     \item \textbf{RQ3: }What is the qualitative performance of MM-SS?
% \end{itemize}

     \begin{table*}[ht]

\caption{Helpfulness Prediction results on Lazada-MRHP dataset. "$\dagger$" are from the open-source code in~\cite{ref15}; "$\natural$" are from~\cite{ref15}; "$\lozenge$" is from~\cite{ref17}; "$\ast$" denotes that the p-value of the significant test between our result and the best baseline result is less than 0.01.}
\centering
    \begin{tabular}{|l|l|ccc|ccc|ccc|}
    \toprule
         \multirow{2}{*}{\textbf{Type}}& \multirow{2}{*}{\textbf{Method}}&\multicolumn{3}{c|}{\textbf{CS\&J}}&\multicolumn{3}{c}{\textbf{Elec.}}&\multicolumn{3}{|c|}{\textbf{H\&K}}  \\
         &&\textbf{MAP}&\textbf{N@3}&\textbf{N@5}&\textbf{MAP}&\textbf{N@3}&\textbf{N@5}&\textbf{MAP}&\textbf{N@3}&\textbf{N@5}\\
         \midrule
         \multirow{4}{*}{Text-only} &BiMPM$^\dagger$ & 60.3 & 51.8 & 57.6 & 74.6 & 67.0 & 71.8 & 70.7 & 64.5 & 69.3 \\
&EG-CNN$^\dagger$ & 60.0 & 51.4 & 57.2 & 73.1 & 66.5 & 71.0 & 70.2 & 63.2 & 68.8 \\
&Conv-KNRM$^\dagger$ & 61.8 & 54.6 & 59.7 & 74.5 & 67.1 & 71.6 & 71.3 & 65.6 & 70.1 \\
&PRH-Net$^\dagger$ & 62.0 & 54.7 & 59.5 & 74.5 & 67.3 & 72.0 & 71.4 & 65.5 & 69.5\\
         \midrule
         \multirow{5}{*}{Multimodal}&SSE-Cross$^\natural$ & 66.1 & 59.7 & 64.8 & 76.0 & 68.9 & 73.8 & 72.2 & 66.0 & 71.0 \\
&DR-Net$^\natural$ & 66.5 & 60.7 & 65.3 & 76.1 & 69.2 & 74.0 & 72.4 & 66.3 & 71.4 \\
&MCR$^\dagger$ & 68.6 & 62.1 & 66.9 & 76.7 & 70.5 & 74.8 & 73.8 & 67.1 & 72.5 \\
&MCL$^\lozenge$ & 70.3 & 64.7 & 69.0 & 78.2 & 72.4 & 76.5 & 75.2 & 68.8 & 73.7 \\
&\textbf{MM-SS}(Ours) & \textbf{71.7$^*$} & \textbf{65.9$^*$} & \textbf{70.4$^*$} & \textbf{79.4$^*$} & \textbf{73.7$^*$} & \textbf{77.7$^*$} & \textbf{76.6$^*$} & \textbf{69.9$^*$} & \textbf{75.0$^*$}\\
         \bottomrule
    \end{tabular}
    \label{tab:label3}
\end{table*}

\begin{table*}[ht]
\caption{Helpfulness Prediction results on Amazon-MRHP dataset. "$\dagger$" are from the open-source code in~\cite{ref15}; "$\natural$" are from~\cite{ref15}; "$\lozenge$" is from~\cite{ref17}; "$\ast$" denotes that the p-value of the significant test between our result and the best baseline result is less than 0.01.}
\centering
    \begin{tabular}{|l|l|ccc|ccc|ccc|}
    \toprule
         \multirow{2}{*}{\textbf{Type}}& \multirow{2}{*}{\textbf{Method}}&\multicolumn{3}{c|}{\textbf{CS\&J}}&\multicolumn{3}{c}{\textbf{Elec.}}&\multicolumn{3}{|c|}{\textbf{H\&K}}  \\
         &&\textbf{MAP}&\textbf{N@3}&\textbf{N@5}&\textbf{MAP}&\textbf{N@3}&\textbf{N@5}&\textbf{MAP}&\textbf{N@3}&\textbf{N@5}\\
         \midrule
         \multirow{4}{*}{Text-only} &BiMPM$^\dagger$ & 58.1 & 41.6 & 46.3 & 52.5 & 40.3 & 44.5 & 56.94 & 43.71 & 47.82 \\
&EG-CNN$^\dagger$ & 56.5 & 40.8 & 44.9 & 51.7 & 39.7 & 42.4 & 55.6 & 42.8 & 47.1 \\
&Conv-KNRM$^\dagger$ & 56.8 & 40.9 & 45.3 & 52.4 & 40.3 & 44.0 & 57.1 & 44.2 & 48.3 \\
&PRH-Net$^\dagger$ & 58.1 & 41.9 & 46.2 & 52.1 & 39.8 & 43.6 & 56.8 & 44.1 & 47.8\\
         \midrule
         \multirow{5}{*}{Multimodal}&SSE-Cross$^\natural$ & 65.0 & 56.0 & 59.1 & 53.7 & 43.8 & 47.2 & 60.8 & 51.0 & 54.0 \\
&DR-Net$^\natural$ & 65.2 & 56.1 & 59.2 & 53.9 & 44.2 & 47.5 & 61.2 & 51.8 & 54.6 \\
&MCR$^\dagger$ & 66.6 & 57.5 & 60.4 & 54.6 & 45.1 & 48.3 & 62.8 & 53.7 & 56.9 \\
&MCL$^\lozenge$ & 67.4 & 58.6 & 61.6 & 56.5 & 47.6 & 50.8 & 63.5 & 54.6 & 57.8 \\
&\textbf{MM-SS}(Ours) & \textbf{69.5$^*$} & \textbf{60.1$^*$} & \textbf{63.3$^*$} & \textbf{58.3$^*$} & \textbf{49.2$^*$} & \textbf{52.5$^*$} & \textbf{65.3$^*$} & \textbf{56.3$^*$} & \textbf{59.4$^*$}\\
         \bottomrule
    \end{tabular}
    \label{tab:label4}
\end{table*}

\subsection{ Experimental Settings}
% \subsubsection{Datasets.} 
We evaluate our model on two public benchmark datasets for the MRHP task: Lazada-MRHP~\cite{ref15} and Amazon-MRHP~\cite{ref15}. Next, we give a brief introduction to the two datasets. (1) \textbf{Lazada-MRHP}  consists of product items and artificial reviews mined from Lazada~\footnote{\url{https://www.lazada.cn}.}, an e-commerce platform in Southeast Asia.  Product information and associated reviews are in Indonesian and extracted between 2018 and 2019. (2) \textbf{Amazon-MRHP}  is collected from Amazon\footnote{\url{https://Amazon.com}.} which is a large-scale international e-commerce platform. Product information and associated reviews are in English and extracted between 2016 and 2018. 

Both datasets comprise 3 categories: (i) Clothing, Shoes $\&$ Jewelry~(CS\&J), (ii) Electronics~(Elec.), and (iii) Home $\&$ Kitchen (H\&K).We present statistics information of datasets in Table~\ref{tab:label1}. All the implementation details can be found in the Appendix.

\begin{table}[ht]
\caption{Statistics of the two datasets. \#P and \#R represent the number of products and reviews.}
    \centering
    \resizebox{\linewidth}{!}{
    \begin{tabular}{l|l|ccc}
    \toprule
         \multirow{2}{*}{\textbf{Dataset}}&\multirow{2}{*}{\textbf{Cate}}&\multicolumn{3}{c}{\textbf{Instance Number(\#P/\#R)}}  \\
         &&\textbf{Train}&\textbf{Dev}&\textbf{Test}\\
         \midrule
         \multirow{3}{*}{Lazada}&CS\&J  &6,598/104,121&1,650/26,185&2,063/32,287\\
         &Elec.  &3,849/41,862&963/10,569&1,204/12,667\\
         &H\&K&2,941/37,014&736/9,615&920/12,552\\
         \midrule
         \multirow{3}{*}{Amazon}&CS\&J &12,701/277,322&3,203/71,458&3,966/87,492\\
         &Elec.  &10,533/259,953&2,672/64,956&3,327/79,570\\
         &H\&K&14,570/369,518&3,616/92,707&4,529/111,193\\
         \bottomrule
    \end{tabular}
}
    \label{tab:label1}
\end{table} 

% \subsubsection{Metrics.} 

\subsection{ On Model Comparison}
To exam the validity of our model, we compared our model with several state-of-the-art baselines. Specifically, all the baselines can be categorized into two groups, i.e., text-only and multimodal. The text-only baselines are presented as follows.
\begin{itemize}
    \item \textbf{BiMPM}~\cite{ref10}: The Bilateral Multi-Perspective Matching network ranks two sentences P and Q by matching sentences in two directions (P against Q and Q against P) with a BiLSTM encoder.
    \item  \textbf{EG-CNN}~\cite{ref11}: The Embedding-gated CNN model applies word-level embedding-gates and cross-domain relationship learning to the review text to predict the review helpfulness score. 
    \item \textbf{ConvKNRM}~\cite{ref31}: The Convolutional Kernel-based Neural Ranking model uses the n-gram soft matches by the kernel pooling to generate the final ranking score. 
    \item \textbf{PRHNet}~\cite{ref5}: The Product-aware Helpfulness Prediction Network is an end-to-end deep neural architecture directly fed by both the metadata of a product and the raw text of its reviews to acquire product-aware review representations for helpfulness prediction.
\end{itemize}

In multimodal settings, we pick a collection of state-of-the-art multimodal helpfulness prediction models for comparison as follows.
\begin{itemize}
\item \textbf{SSE-Cross}~\cite{ref6}: The Stochastic Shared Embeddings Cross-modal Attention Network is a multimodal framework which trains image and text pairs with stochastic shared embeddings.
\item \textbf{D\&R Net}~\cite{ref36}: The Decomposition and Relation Network is a decomposition network that represents the commonality and discrepancy between image and text, and the relation network models the semantic association in cross-modality context.
\item \textbf{MCR}~\cite{ref15}: The Multi-perspective Coherent Reasoning is a product-review coherent reasoning model to measure the intra- and inter-modal coherence between the target product and the review.
\item \textbf{MCL}~\cite{ref17}: The Multi-modal Contrastive Learning concentrates on mutual information between input modalities to explicitly elaborate cross-modal relations.
\end{itemize}

Tables \ref{tab:label3} and \ref{tab:label4} present the results of MM-SS and the baselines on Lazada-MRHP and Amazon-MRHP, respectively. The results yield several key observations: 1) Among all the baselines, EG-CNN performs the worst, suggesting that focusing solely on hidden features extracted from the review text, without considering product information, is not a reasonable approach. 2) Multimodal baselines consistently outperform those that rely solely on textual data (i.e., BiMPM, EG-CNN, Conv-KNRM, and PHRNet). This underscores the value of incorporating both textual and visual modalities in the model, as it enables the extraction of more helpful information. 3) MM-SS consistently outperforms all baselines across various metrics on both datasets. This highlights the advantage of our approach, which leverages self-supervised multi-task learning. And 4) It's noteworthy that our approach exhibits a substantial performance improvement on Amazon-MRHP compared to Lazada-MRHP. We speculate that this disparity arises from the better alignment of our fine-tuned BERT model with the English context, as opposed to Indonesian. 

It's worth noting that we conducted a significance test between our results and the best baseline result (i.e., MCL) to establish the statistical significance of our improvements. All the p-values were found to be less than 0.01, further substantiating the superiority of our method over existing approaches.

\subsection{ On Ablation Study}
To verify the importance of each component in our model, we also compare MM-SS with the following derivatives.
\begin{itemize}
\item \textbf{w/o intra-modal-$p_{t}r_{t}$:} To explore the effect of the intra-modal interactive subtask between product and review text, we remove this component by setting $s\in\{p_{v}r_{v},p_{t}r_{v},p_{v}r_{t}, r_{t}r_{v}\}$ in section \ref{section:D}.
\item \textbf{w/o intra-modal-$p_{v}r_{v}$:} To explore the effect of the intra-modal interactive subtask between product and review image, we remove this component by setting $s\in\{p_{t}r_{t},p_{t}r_{v},p_{v}r_{t}, r_{t}r_{v}\}$ in section \ref{section:D}.
\item \textbf{w/o inter-modal-$p_{t}r_{v}$:} To explore the effect of the inter-modal interactive subtask between product text and review image, we remove this component by setting $s\in\{p_{t}r_{t},p_{v}r_{v},p_{v}r_{t}, r_{t}r_{v}\}$ in section \ref{section:D}.
\item \textbf{w/o inter-modal-$p_{v}r_{t}$:} To explore the effect of the inter-modal interactive subtask between product image and review text, we remove this component by setting $s\in\{p_{t}r_{t},p_{v}r_{v},p_{t}r_{v}, r_{t}r_{v}\}$ in section \ref{section:D}.
\item \textbf{w/o intra-review-$r_{t}r_{v}$:} To explore the effect of the intra-review interactive subtask between review text and review image, we remove this component by setting $s\in\{p_{t}r_{t},p_{v}r_{v},p_{t}r_{v}, p_{v}r_{t}\}$ in section \ref{section:D}.
\item \textbf{w/o Modality-specific Feature Extraction (module b):} To gain more insights into our manners of utilizing textual and visual information, we directly concatenated the textual and visual features of each item and fed them to a MLP to get $\hat{y}_g$. Accordingly, the Pseudo-Label Generation Subtasks and Global Feature Fusion Target-task are simultaneously removed.
\item \textbf{w/o Fine-grained Contribution-guided Helpfulness Measurement (module d):}  We remove all five subtasks to further investigate whether the improvement of MM-SS is achieved by self-supervised pseudo-label generation scheme.
\end{itemize}

Due to space constraints, we conducted the ablation study on the CS\&J category of Lazada and Amazon datasets, with the results presented in Table~\ref{tab:label5}. The table reveals the following insights:

1) In all subtasks, the  intra-review-$r_{t}r_{i}$ task has the most significant impact on the performance of MM-SS. This suggests that images within a review serve as valuable evidence for review helpfulness prediction. However,  intra-modal-$p_{t}r_{t}$ and intra-modal-$p_{v}r_{v}$ have a smaller influence on MM-SS compared to the other two variants, inter-modal interaction $p_{v}r_{t}$ and $p_{t}r_{v}$. This could be attributed to most product images being consistently enhanced, resulting in substantial distinctions between product images and images shared by consumers.
2) w/o Fine-grained Contribution-guided Helpfulness Measurement module yields poorer performance. For instance, in the Lazada-MRHP dataset, MAP decreases by 1.97 points, and N@5 decreases by 1.56 points. A similar trend is observed in the Amazon-MRHP dataset, which underscores the effectiveness of our proposed Pseudo-Label Generation scheme.
3) MM-SS outperforms the w/o Modality-specific Feature Extraction module. For example, in the Amazon-MRHP dataset, MAP decreases by 5.57 points, and N@5 decreases by 5.91 points. This pattern is also observed in the Lazada-MRHP dataset. It is hypothesized that without this module, the model rigidly relies on the alignment between multimodal input elements, resulting in an impractical model because, in most cases, cross-modal elements cannot be accurately mapped together\cite{ref16}.

 \begin{table}[ht]
 \caption{ The ablation study on CS\&J category of Lazada-MRHP and Amazon-MRHP.}
    \centering
    \resizebox{\linewidth}{!}{
    \begin{tabular}{l|l|ccc}
    \toprule \textbf{Dataset} & \textbf{Model Variant} & \textbf{MAP} & \textbf{N@3} & \textbf{N@5} \\
    \midrule \multirow{9}{*}{ Lazada } & \textbf{MM-SS} (Ours) & \textbf{71.73} & \textbf{65.94} & \textbf{70.42} \\
    & -w/o intra-modal-$p_{t}r_{t}$ & 71.41 & 65.85 & 70.20 \\
    & -w/o intra-modal-$p_{v}r_{v}$ & 71.29 & 65.73 & 70.15 \\
    & -w/o  inter-modal-$p_{t}r_{v}$ & 70.94 & 65.52 & 69.93 \\
    & -w/o  inter-modal-$p_{v}r_{t}$ & 71.06 & 65.51 & 70.08 \\
    & -w/o intra-review-$r_{t}r_{v}$ & 70.86 & 65.43 & 69.87 \\
    & -w/o module(b) & 65.34 & 60.25 & 64.93 \\
    & -w/o module(d) & 69.76 & 64.37 & 68.86 \\
    \midrule \multirow{9}{*}{ Amazon } & \textbf{MM-SS} (Ours) &\textbf{69.54} & \textbf{60.12} &   \textbf{62.94} \\
    & -w/o intra-modal-$p_{t}r_{t}$ & 69.04 & 60.05 & 62.88 \\
    & -w/o intra-modal-$p_{v}r_{v}$ & 68.86 & 59.97 & 62.75 \\
    & -w/o  inter-modal-$p_{t}r_{v}$ & 68.63 & 59.66 & 62.47 \\
    & -w/o  inter-modal-$p_{v}r_{t}$ & 68.65 & 59.70 & 62.36 \\
    & -w/o intra-review-$r_{t}r_{v}$ & 68.46 & 59.38 & 62.24 \\
    & -w/o module(b) & 63.97 & 54.81 & 57.03 \\
    & -w/o module(d) & 68.02 & 58.93 & 61.47 \\
    \bottomrule
\end{tabular}
} 
    \label{tab:label5}
\end{table}

\subsection{ On Case Study}
\begin{figure}[ht]
    \centering
    \includegraphics[width=\linewidth]{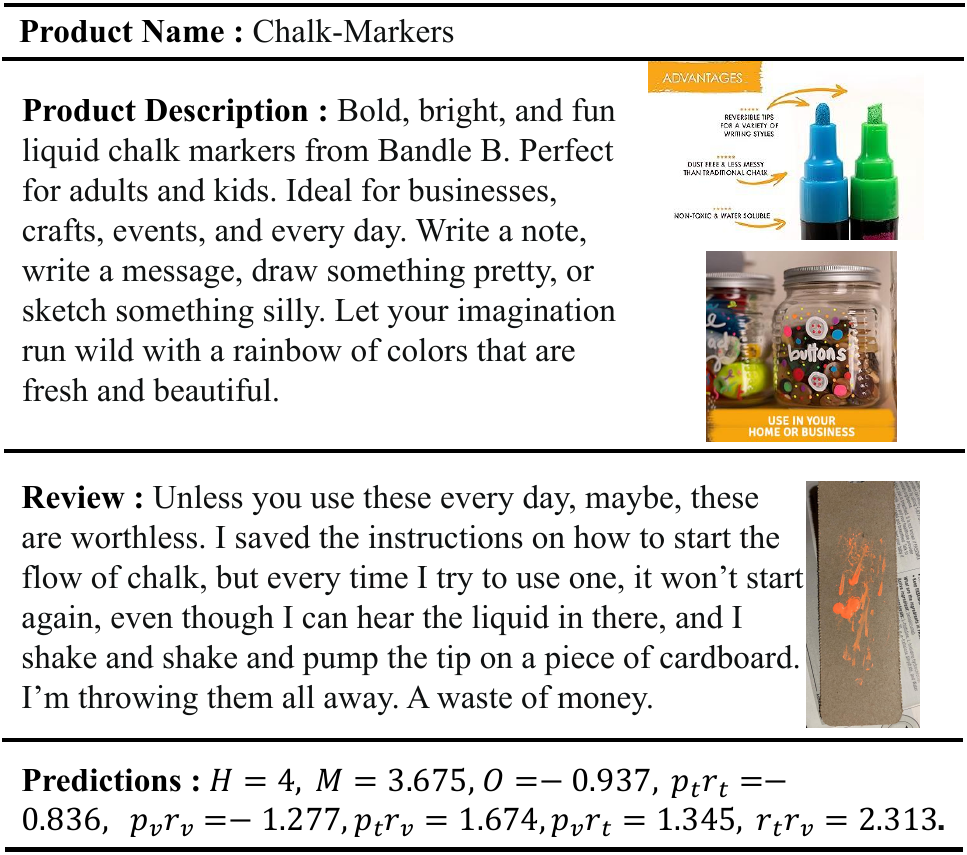}
    \caption{Case study for the MM-SS on Amazon-MRHP. The "H" is human-annotated, "M" is our model prediction, "O" is our model prediction without SSP-Labels, and others are auto-generated SSP-Labels.  }
      \label{fig:label4}
\end{figure}

To validate the rationality of auto-generated Pseudo-labels, we select a multimodal example from the test set of Amazon Home$\&$Kitchen, as shown in Figure \ref{fig:label4}. In this case, manual-annotate label is $4.0$, our model prediction is $3.675$, our model prediction without SSP-Labels is $-0.937$. In line with expectation, the SSP-Label outputs a positive offset, helping us achieve results close to the manual-annotate label. So we believe that these independent pesudo labels can aid in learning both modal consistency and differentiation.

\section{Conclusion}
\label{sec:conclusion}
%In review ranking scenarios, the target task can be improved via training subtasks on a multi-task learning network. In this work, we introduce separate cross-modal interaction subtasks to aid in modeling the consistent and different relations between the visual and textual modalities for the MRHP target task. Different from previous works, we design a pseudo label generation strategy based on the self-supervised method, which saves lots of labor- and time-cost. Extensive experiments validate the rationality of the auto-generated pseudo labels. Our framework can outperform prior textual and multimodal baselines on two real-world datasets. We hope this work can provide a new perspective on the MRHP task.

%We also find that the generated $y_{p_tr_t}$ and $y_{p_vr_v}$ labels are not significant enough limited by the pre-processed features. Future research can be focused on the following aspects: 1) we will explore the connotative relationship between multimodal interactions to find more valuable subtasks, 2) we will fine-tune BERT so that it can be generalized to more language models, 3)in addition to deep learning methods, we consider integrating syntactic features used in traditional machine learning feature engineering methods, such as sentiment analysis \cite{ref54,ref55}.
In this work, we introduce separate cross-modal interaction subtasks to aid in modeling the consistent and different relations between the visual and textual modalities for the MRHP target task. Extensive experiments validate the rationality of the auto-generated pseudo labels. Our framework can outperform prior textual and multimodal baselines on two real-world datasets. We also find that the generated $y_{p_tr_t}$ and $y_{p_vr_v}$ labels are not significant enough limited by the pre-processed features.
Future research can be focused on the following aspects: 1) we will explore the connotative relationship between multimodal interactions to find more valuable subtasks, 2) we will fine-tune BERT so that it can be generalized to more language models, 3)in addition to deep learning methods, we consider integrating syntactic features used in traditional machine learning feature engineering methods, such as sentiment analysis \cite{ref54,ref55}.

\section{Conclusion}\label{sec:conclude}
In this work, we tackle review helpfulness prediction using two new techniques, i.e., embedding-gated CNN and cross-domain relationship learning. We built our base model on CNN with word-, character- and topic-based representations.
On top of this model, 
domain relationships were learned to better transfer knowledge across domains. 
%Inspired by the fact that different words may play roles in idea expression, we proposed an embedding-gated CNN which shows significantly better results than the competing methods. We further improved our model by considering cross-domain relationship learning.
The experiments showed that our model significantly outperforms the state of the art. % and the domain relationships learned are insightful.% and helpful.% for the task. 
%Our model is a neural network based model without using handcrafted linguistic features. In the near future, we seek to incorporate more linguistic features to examine their importance to the task.

%Review helpfulness prediction has been an important module for E-commerce websites. Recent studies on review helpfulness rely on sufficient labeled samples with hand-crafted features for product domains/categories of interests. In this work, we introduce a deep convolutional neural network model that is able to extract high-level features from character, word and topic-based representations. Inspired by the fact that different words may play different weights in the final results, we consider to learn embedding-gates for word representations.
%Extensive evaluations show our model yields better performance than the existing methods. We further integrate our model into a cross-domain relationship learning framework to jointly model different domains to help domain of interests. Our experiments show that the learnt domain relationship is insightful and helpful for our task.

\bibliographystyle{ieeenat_fullname}
\bibliography{ijcai18}

\clearpage
\appendix
\section{Appendix}
\subsection{Offset Value} 
Firstly, in the training process,we abstract an anchor point: 
\begin{equation} 
\textbf{A}_{k}=\frac{\sum\nolimits_{i=1}^N \sigma(y_{k}(i))\cdot \epsilon_{k}^{i}}{\sum\nolimits_{i=1}^N \sigma(y_{k}(i))},
\end{equation}
where $k\in\{g,p_{t}r_{t},p_{v}r_{v},p_{t}r_{v},p_{v}r_{t},r_{t}r_{v}\}$, $N$ is the number of training samples, $\epsilon_{k}^j$ is the modification value of the $i$-th sample in task $k$. $\sigma$ is the Sigmoid function. As modal representations reside in distinct feature spaces, it is inaccurate to solely employ Euclidean distances for their measurement. Consequently, we employ Mahalanobis Distance~\cite{ref53} to normalize them, ensuring an equal distribution of dimensions. The distance from representations to the anchor is calculated as follows

\begin{equation} 
\chi_{k}=\frac {\sqrt{(f_{k}^{*}-\textbf{A}_{k})^{T}\sum^{-1}(f_{k}^{*}-\textbf{A}_{k})}} {\sqrt{s_{k}}},
\end{equation}
where $s_k$ is a scale factor, $\Sigma^{-1}$ is the covariance matrix between $f_{k}^*$ and anchor. Then we calculate the distance difference from representation to the anchor.

According to the correlation analysis $\frac {{y}_{s}} {{y}_{g}}\propto \frac {{\chi}_{s}} {{\chi}_{g}}$ in Chapter \ref{section:D}, we can make the following transformation to the formula,
\begin{equation}
y_s =  (\alpha_1 * \frac{\chi_s}{\chi_g+\epsilon}) * y_g,
\end{equation}
where $s\in\{p_{t}r_{t},p_{v}r_{v},p_{t}r_{v},p_{v}r_{t}, r_{t}r_{v}\}$, $\alpha_1$ is the correlation coefficient, $\epsilon$ is a minuscule value utilized to avoid division by zero in the denominator. 

From a subtraction perspective, we can infer the following $y_{s}-y_{g} \propto \chi_{s}-\chi_{g}$. Then we approach the formula for further modifications,
\begin{equation}
\begin{aligned}
    y_{s}=y_{g}+\alpha_2 * (rel_{s}-rel_{g}),
\end{aligned}
\end{equation}
where $\alpha_2$ is the correlation coefficient.

By combining the above two equations and assigning equal weights, we can calculate the offset,
\begin{equation}
\label{con:inventoryflow}
\begin{aligned}
    y_{s}= &\frac{y_g + \alpha_2 * (\chi_s - \chi_g)}{2} + \frac{\alpha_1*\chi_s y_g}{2(\chi_g+\epsilon)} \\
=& y_{g}+ \frac{2\chi_s-\alpha*(\chi_g+\epsilon)y_g}{2(\chi_g+\epsilon)} \\
=& y_{g}+\delta_{gs},
\end{aligned}
\end{equation}
where $\delta_{gs}$ denotes the offset value of subtask supervisions from the target-task output.

\subsection{Implementation Details} 
As for the text encoder, we set the word embedding size to $512$, and the dimension of the hidden layer in BERT~\cite{ref27} to $128$. Regarding the image encoder, we extract 2048-dimensional ROI features from each image and encode them into 128-dimensional vectors. Our model is trained end-to-end with Adam optimizer \cite{ref35} and the batch size is set to $32$. We use the initial learning rate of $5e\text{-}5$ for BERT and $1e\text{-}4$ for other parameters. For the training objective, we set the value of the margin in the ranking loss to $1$. To make a fair comparison, we run five times and report the average result of all the models. All the experiments are implemented by PyTorch\footnote{\url{https://pytorch.org}.} over a server equipped with $1$ NVIDIA GTX-1080Ti GPU, and the random seeds are fixed for reproducibility. 

Following previous MRHP studies~\cite{ref15,ref16,ref17}, we adopt several retrieval metrics for the evaluation of models, i.e., Mean Average Precision~(MAP) and Normalized Discounted Cumulative Gain~(NDCG@N)~\cite{ref33,ref34}, and we set $N$ as $3$ and $5$, respectively. For comparison, we rank all reviews in descending order by helpfulness score. MAP reflects the average accuracy of the entire set of reviews.  
Besides, NDCG@N evaluates the performance of generating the top $N$ helpful review lists for different customers, which simulates a scenario where customers would only read a limited few reviews.
\end{document}